\documentclass[reprint,nofootinbib,prd,aps]{revtex4-2}

\usepackage{graphicx}
\usepackage{dcolumn}
\usepackage{bm}
\usepackage{caption}
\usepackage{amsmath,amssymb}

\usepackage{color}
\usepackage[dvipsnames]{xcolor}

\usepackage{aas_macros}
\usepackage{hyperref}

\newcommand{\ag}{\alpha_\text{grav}}

\begin{document}


\title{Improved constraints on modified Newtonian gravity \\ from Cassini radio tracking data}

\author{R. S. Park}
\affiliation{%
 Jet Propulsion Laboratory, California Institute of Technology, 4800 Oak Grove Drive, Pasadena, CA 91109, USA\\
}%

\author{A. Hees}
\affiliation{LTE, Observatoire de Paris, Universit\'e PSL, Sorbonne Universit\'e, Universit\'e de Lille, LNE, CNRS 61 Avenue de l'Observatoire, 75014 Paris, France\\
}%

\author{B. Famaey}
\affiliation{%
 Université de Strasbourg, CNRS UMR 7550, Observatoire astronomique de Strasbourg, 11 rue de l’Université, 67000 Strasbourg, France
}%

\author{H. Desmond}
\affiliation{
Institute of Cosmology \& Gravitation, University of Portsmouth, Dennis Sciama Building, Portsmouth, PO1 3FX, UK
}%

\author{A. Durakovic}
\affiliation{
University of Sarajevo (UNSA), Faculty of Science, Zmaja od Bosne 33-35, 71000 Sarajevo, Bosnia and Herzegovina
}%

\date{\today}

\begin{abstract}
We report an updated constraint on the Solar System quadrupole parameter $Q_2$, which encodes the external field effect predicted by modified gravity versions of the MOND paradigm. Using the dataset employed to compute the DE440 planetary ephemerides, and estimating it simultaneously with other parameters included in the planetary ephemerides, we find $Q_2 = (1.6 \pm 1.8) \times 10^{-27}\,\mathrm{s}^{-2}$ (1-$\sigma$), representing an improvement of 40\% over previous estimates. We also show explicitly that the contribution to the MOND prediction of $Q_2$ from the Solar System’s largest planet, Jupiter, is at the 0.05\% level, validating the approximation of retaining only the Sun in theoretical calculations. With this new constraint on $Q_2$, we update previously acknowledged tensions with external galaxy rotation curves, now leading to discrepancies at the $3$-$15\sigma$ level depending on the detailed mass modeling or the subset of galaxies considered. Within the Milky Way itself, the $Q_2$ constraint imposes an upper bound of only 2\% (at 95\% confidence) on the MOND boost to the galactic radial acceleration (i.e., the ratio of the observed over baryonic Newtonian acceleration) at the position of the Sun, in strong tension with current observational limits. The updated $Q_2$ posterior finally confirms that Solar System measurements provide stronger constraints than current wide-binary data on classical modified gravity versions of MOND. 

\end{abstract}

\maketitle

\section{Introduction}

The Dark Matter (DM) problem, which arises from discrepancies between observations of galactic and extragalactic systems (including the Universe as a whole) and the predictions of General Relativity combined with the Standard Model of particle physics, is one of the central problems of modern physics, since the nature of DM has remained elusive for several decades. This DM, or something behaving like it, is hard to avoid cosmologically, and has therefore been considered for the last five decades as forming invisible halos within which galaxies form \citep[e.g.,][]{White}. This hypothesis supposedly explains the observations of disk galaxy rotation curves deviating from the quasi-Keplerian decline expected in the presence of only baryons \citep[e.g.,][]{Bosma, Rubin}. However, a possible alternative in galaxies is that dynamics actually deviates from Newtonian expectations in the ultra-weak acceleration regime. This hypothesis, known as modified Newtonian dynamics \citep[MOND,][]{Milgrom83}, has proven surprisingly successful in explaining a large range of observations at galaxy scales \citep[e.g.,][]{FamaeyMcG, FamaeyDura}.

The simplicity and regularity of observed disk galaxy kinematics is epitomized by the radial acceleration relation \citep[RAR,][]{McGaugh,Lelli}, which precisely relates the total gravitational acceleration at any point in a galactic disk to that generated by the observed baryons, as was predicted by Milgrom \citep{Milgrom83} before the relation was empirically established. The RAR exhibits remarkably small intrinsic scatter and lack of residual correlations with other galaxy properties, and it has been shown to be the root cause of all other scaling relations in the dynamics of disk galaxies \citep{Stiskalek}, and of the observed diversity of inner shapes of rotation curves, strongly correlated to the baryonic surface density even in galaxies where dark matter is supposed to dominate the inner regions \citep[e.g.,][]{deBlok, Swaters, Ghari}. This is not fully understood in the presence of DM particles that would interact with baryons only through gravity, and may indicate that dynamics or gravity is effectively modified at galaxy scales, or that DM interacts with baryons in more subtle ways than usually considered.

The MOND paradigm introduces a new fundamental constant of acceleration, $a_0 \sim 10^{-10} \, {\rm m}{\rm s}^{-2}$, such that, when accelerations $a$ are well below this threshold, \emph{either} the force of inertia on a mass $m_1$ is modified, and goes from $m_1 a$ to $m_1 a^2/a_0$ (for circular motion at least) \cite{inertia}, \emph{or} the gravitational force between two point masses $m_1$ and $m_2$ separated by a distance $r$ (and isolated from any third body) is modified and goes from $G m_1 m_2/r^2$ to $2\sqrt{Ga_0}\left[ (m_1+m_2)^{3/2} - m_1^{3/2} - m_2^{3/2} \right]/3r $. In the case where $m_1 \ll m_2$, this two-body force becomes $m_1 \sqrt{Gm_2a_0}/r$ \citep[see, e.g.,][]{FamaeyDura}, which thereby is straightforwardly equivalent to the first formulation for a test particle moving in the gravitational field generated by $m_2$. The first option is known as modified inertia, and is hard to implement in a covariant framework extending General Relativity. Therefore, it has never been theoretically fully developed beyond nonrelativistic toy models \citep{inertia}. The second option, modified gravity, has on the other hand led to several covariant relativistic formulations \citep[][]{Skordis,BIMOND,Blanchet,Deffayet} that have been developed over the years, and are still under development. In the non-relativistic limit relevant for the Solar System, these modified gravity approaches mostly \citep[but see, e.g.,][for details]{FamaeyDura} reduce to two variations: (i) the aquadratic Lagrangian (AQUAL) where the weak-field gravitational potential obeys a non-linear Poisson equation akin to the one of the electric field inside a dielectric medium, and (ii) the quasilinear MOND (QUMOND) formulation where it obeys a linear Poisson equation, but sourced by a virtual source term (including what is called ``phantom'' dark matter) non-linearly related to the baryonic distribution. Any such modification of gravity must however pass the tight gravity tests that can be conducted in the Solar System. 

Within the Solar System, three types of effects can arise due to MOND as modified gravity. All three effects are related to the shape of the transition between the deep-MOND regime ($a \ll a_0$) and the Newtonian regime ($a \gg a_0$), parametrized by the so-called interpolating function (IF) of MOND. The first effect is generic (i.e., it belongs in principle to modified inertia as well as modified gravity), and is related to how fast the gravitational field transitions to the Newtonian one, \emph{above} $a_0$. This is equivalent to quantifying the residual anomalous force at high accelerations. Some forms of the IF that account well for the dynamics of galaxies can be excluded solely from this residual anomalous force  \cite{sereno:2006vn}, but this can be easily cured so that the departure from unity of the IF, meaning departure from Newtonian dynamics, becomes fully negligible for gravitational fields relevant to the Solar System \citep[see, e.g.,][]{FamaeyMcG}. The second effect pertains to the fact that the Solar System is actually aspherical. In an aspherical system, even an  isolated one, an anomalous quadrupolar contribution to the potential generically appears in modified gravity MOND \citep{Milgrom2012}, with a strength proportional to $(a^N/a_0)^{-5/2}$, where $a^N$ is the Newtonian gravitational acceleration. Again, if the IF approaches Newtonian dynamics fast enough \emph{above} $a_0$, the effect can be made hardly detectable in the inner Solar System \cite{Milgrom2012,iorio:2013ty}. The third effect is the most stringent one and is related to the so-called external field effect (EFE) of MOND. The EFE implies that the internal gravitational dynamics of a system depends on the external gravitational field in which the system is embedded \cite{bekenstein:1984kx} and therefore breaks the strong equivalence principle. In the inner Solar System, this effect mostly manifests itself as a quadrupole modification to the Newtonian potential \cite{milgrom:2009vn}. This quadrupole field is a solution of the Laplace equation, hence associated to zero phantom density locally, but is affected by the asymmetric distribution of the MOND phantom density at large distances, where the external field from the Milky Way and the internal field of the Solar System have about the same magnitude. In the axisymmetric approximation, the quadrupole field is typically quantified by one parameter for its amplitude, noted $Q_2$.

In \cite{Desmond_Cassini}, it has been explicitly shown that the predicted value of $Q_2$ depends solely on the behavior of the IF at the value of the external gravitational field acting on the Sun, hence close to $a_0$. This means that making the IF approach Newton as fast as possible \emph{above} the Milky Way gravitational field acting on the Sun does \emph{not} decrease the value of $Q_2$, making its measurement particularly important and constraining. A first estimate of this $Q_2$ parameter based on nine years of Cassini data has been inferred in \cite{hees_2014}: $Q_2 = \left(3\pm 3\right)\times 10^{-27}$ s$^{-2}$. This result has subsequently been used to constrain the various MOND IF \cite{hees:2016mi,Desmond_Cassini} giving rise to a tension between the Solar System constraint and observations at Galactic scales.

The goal of this manuscript is threefold: (i) clarify that considering only the Sun and not the effect of other planets such as Jupiter on the quadrupole, is sufficient for the targeted accuracy of the $Q_2$ constraint (Sec.~\ref{sec:theory}); (ii) improve the actual $Q_2$ estimate from \cite{hees_2014}, by using the recent planetary ephemerides DE440 \cite{park_2021} that include an improved and complete Cassini dataset (Sec.~\ref{sec:analysis}); (iii) explore the implication of this new observational result when combined to galactic constraints (Sec.~\ref{sec:results}), and show that a precise enough mass modeling of the Milky Way galaxy alone, combined with the $Q_2$ constraint, is largely sufficient to rule out AQUAL and QUMOND (Sec.~\ref{sec:test_MOND_MW}), essentially implying that any MOND-based modified gravity must probably involve another scale beyond $a_0$ \cite{GQUMOND}. In the data analysis (Sec.~\ref{sec:analysis}), we insist that the $Q_2$ parameter is estimated {\it simultaneously} with all other parameters included in the planetary ephemerides. In addition, we will take great care in validating our results, and to search for systematics by estimating this parameter using independent subsets of the data and checking that the estimate is stable.

\section{Theoretical modeling of the MOND quadrupolar signature}
\label{sec:theory}

\subsection{Brief overview of MOND as modified gravity}

As pointed out before, in the non-relativistic weak-field limit relevant to Solar System dynamics, most modified gravity theories of MOND lead to either an AQUAL or a QUMOND limit. In the latter, it is convenient to write the gravitational field as $\bm g=-\bm \nabla \Phi$ where $\Phi=\Phi_N + \Phi_\mathrm{p}$ is expressed as the sum of the Newtonian potential $\Phi_N$, obeying the Newtonian Poisson equation
\begin{equation}
    \nabla^2 \Phi_N = 4 \pi G \rho_\mathrm{b} \, ,
\end{equation}
with $\rho_\mathrm{b}$ the baryonic matter density, and of a potential $\Phi_\mathrm{p}$ solution to
\begin{equation}\label{eq:Phi_p}
    \nabla^2 \Phi_\mathrm{p} = 4 \pi G \rho_\mathrm{p} \,,
\end{equation}
where $\rho_\mathrm{p}$ is an \emph{effective} (virtual) phantom dark matter density defined by
\begin{equation}\label{eq:rho_p}
    \rho_\mathrm{p}=\frac{1}{4\pi G}\bm \nabla\cdot \left[\left(\nu\left(\frac{\left|\bm a^N\right|}{a_0}\right)-1\right)\bm \nabla \Phi_N \right] \, .
\end{equation}
 In the above expression, $\bm a^N=-\bm\nabla\Phi_N$ is the Newtonian gravitational field sourced by baryons, $a_0$ is the MOND acceleration scale and the function $\nu(x)$ is the MOND IF, whose asymptotic behaviors are set by the Newtonian ($\nu(x) \rightarrow 1$ for $x \gg 1$) and MONDian ($\nu(x) \rightarrow x^{-1/2}$ for $x \ll 1$) regimes. In QUMOND, the precise form of this IF is not derived from first principles, hence several functions and classes of functions have been considered in the literature, see e.g. \cite{FamaeyMcG,FamaeyDura}. Here we consider the same IF families as in \cite{Desmond_Cassini}:
\begin{subequations}\label{eq:if_fams}
	\begin{eqnarray}
    \nu_n(x)&=&\left[\frac{1+\left(1+4x^{-n}\right)^{1/2}}{2}\right]^{1/n}\, ,  \label{eq:nun}\\
    \nu_\delta(x)&=&\left(1-e^{-x^{\delta/2}}\right)^{-1/\delta}\, , \label{eq:IF_delta}\\
	\nu_\gamma(x)&=&\left(1-e^{-x^{\gamma/2}}\right)^{-1/\gamma}+\left(1-1/\gamma\right) e^{-x^{\gamma/2}} \, . \label{eq:IF_gamma}
	\end{eqnarray}
\end{subequations}
The parameters $n$, $\delta$ and $\gamma$ characterize the sharpness of the transition between the regimes: larger values correspond to steeper transitions between the Netwonian and MONDian regimes. Limiting cases of these are the well-known Simple ($n=1$) and Standard ($n=2$) IFs, as well as the RAR IF ($\delta=\gamma=1$). The Simple IF ($n=1$) is a prime example of an IF that works remarkably well for rotation curves \cite{FamBin,Gentile} but can readily be excluded in the Solar System \cite{sereno:2006vn} because it predicts a residual anomalous gravitational acceleration of the order of $a_0$, corresponding to a divergent (`cuspy') phantom density around the Sun. On the other hand, the RAR \cite{McGaugh} IF ($\delta=\gamma=1$) is almost indistinguishable from the Simple IF ($n=1$) in galaxies but does not suffer from such a problem in the Solar System.

All computations in this section can be applied similarly within the AQUAL formulation of MOND. The exact expression of the phantom density from Eq.~(\ref{eq:rho_p}) is modified \citep[see, e.g.,][]{Stahl} and the corresponding Poisson equation is actually non-linear \cite{blanchet:2011ys}. As a consequence, the potential needs to be determined numerically. However, the leading effect studied below is predicted to be slightly larger in AQUAL, rendering the QUMOND computations conservative \cite{blanchet:2011ys,Desmond_Cassini}.

\subsection{The QUMOND quadrupole in the Solar System}\label{sec:Q2_solar_system}
\begin{figure*}[htb]
  \centering
  \includegraphics[width=0.8\textwidth]{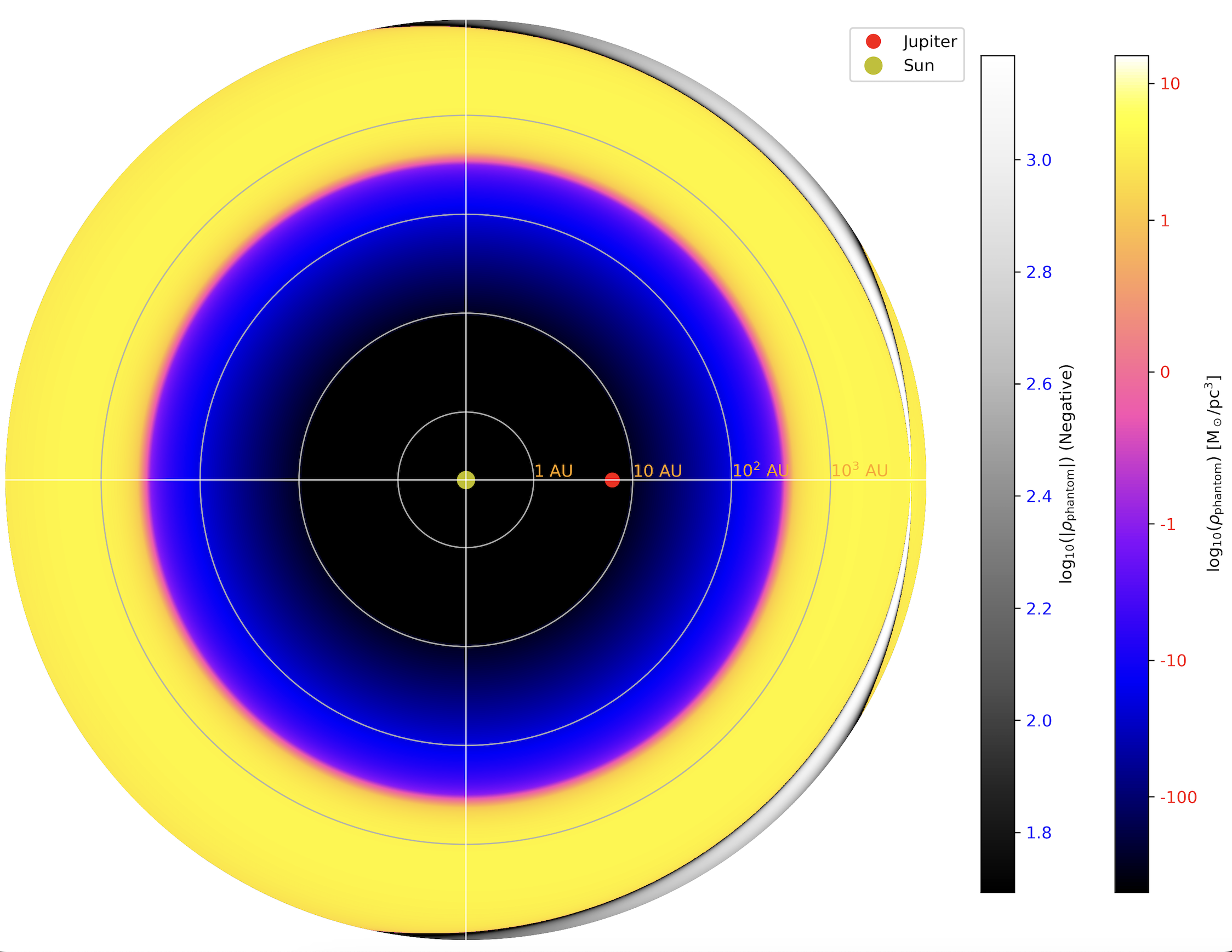}
  \caption{Distribution of the phantom density for the $\delta=\gamma=1$ IF (the RAR IF), computed using Eq.~(\ref{eq:rho_p}) including both the Sun and Jupiter as sources. The Galactic Center is located in the $+z$ direction, i.e. towards the right of the figure. All scales are logarithmic. Clearly, the phantom density is essentially zero within 100~au and the QUMOND quadrupole is entirely related to the aspherical phantom density at much larger distances.}
  \label{fig:phantom}
\end{figure*}

As outlined above, MOND generates three main types of effects in the Solar System, one which is generic but easy to avoid with a sharp enough IF (e.g., $\delta \geq 1$ or $\gamma \geq 1$), the second one is specific to modified gravity but can also be made hardly detectable with a sharp enough IF. The only effect that is hard to avoid in QUMOND and AQUAL is a quadrupole term in the potential that depends only on the value of the IF for the external field of the Milky Way acting on the Sun, hence in the intermediate regime, independently of the sharpness of the transition well above $a_0$. 

Starting from the phantom density distribution around the Solar System given by Eq.~(\ref{eq:rho_p}) with the Sun as the only gravitational source, the general solution to Eq.~(\ref{eq:Phi_p}) is given by
\begin{equation}
\label{eq:exact}
    \Phi_\mathrm{p} = -G\iiint \frac{d^3\bm x'}{\left|\bm x-\bm x'\right|} \rho_\mathrm{p}(\bm x') \, .
\end{equation}

Let us now consider an IF that is almost exactly Newtonian in the inner Solar System (i.e., $\nu = 1$), such that $\Phi_\mathrm{p}$ is locally solution to the Laplace equation, since there is essentially no phantom density locally, and can be written as a symmetric trace free multipolar expansion \cite{milgrom:2009vn}. The first dipolar term vanishes in a reference system following the barycenter. Therefore, the leading MOND anomalous contribution takes the form of a symmetric trace free quadrupole, i.e. 
\begin{equation}\label{eq:Phi_EFE}
\Phi_\mathrm{p}(\bm x) \simeq \delta \Phi(\bm x) \equiv - \frac{Q_2}{2} \left( \left(\hat {\bm e}\cdot \bm x\right)^2 - \frac{\left|\bm x\right|^2}{3} \right).
\end{equation}
Here, $\hat {\bm e}$ is a unit vector pointing in the direction of the Galactic external field, hence in the direction of the Galactic Center, and $Q_2$ is the amplitude of the quadrupole. 

We can then substitute Eq.~(\ref{eq:Phi_EFE}) into the left-hand side of Eq.~(\ref{eq:exact}), and Eq.~\eqref{eq:rho_p} into the right-hand side of Eq.~(\ref{eq:exact}). Taking the directional second derivative of the phantom potential along $\hat {\bm e}$ at the center then allows one to obtain the analytical formulation of $Q_2$  (see Eq.~23 from \cite{milgrom:2009vn}) as\footnote{In his original paper \cite{milgrom:2009vn}, Milgrom rather used $q_{zz}^a = -(2/3)Q_2$.}
\begin{equation}\label{eq:q_general}
    Q_2=\frac{3}{8\pi}\iiint \mathrm{d}^3\bm x \frac{\left(\nu -1\right)}{\left|\bm x\right|^4} \Big[6a_z^N n_z+3\bm a^N \cdot \bm n (1-5n_z^2)\Big] \, ,
\end{equation}
where $\bm n=\bm x/\left|\bm x\right|$ and where the $z$ direction is assumed to be aligned with the external Galactic field.

If we model the Solar System by considering only a spherically symmetric Sun as the internal source of the gravitational field, the total Newtonian acceleration appearing in Eq.~(\ref{eq:q_general}) is
\begin{equation}\label{eq:gN}
    \bm a^N = \bm a^N_e - \frac{GM_\odot}{\left|\bm x\right|^3}\bm x = a_0 \left(e_N \hat{\bm e} - \frac{\bm u}{\left|\bm u\right|^3} \right)\, ,
\end{equation}
where $\bm u=\bm x/R_M$, $e_N=\left|\bm a_{e}^N\right|/a_0$, with $R_M=\left(GM_\odot / a_0\right)^{1/2}$, the MOND radius, and $\bm a_e^N$ the Newtonian external Galactic field. Inserting this expression in Eq.~(\ref{eq:q_general}) leads to the usual quadrupole formula
\begin{subequations}\label{eq:Q2_sun_both}
\begin{equation}\label{eq:Q2_Sun}
    Q_2 = -\frac{3a_0}{2R_M}q=-\frac{3a_0^{3/2}}{2\sqrt{GM_\odot}}q \, ,
\end{equation}
 with $q$ a dimensionless quadrupole that depends only on the MOND IF and on the Newtonian value of the external field, whose expression in the case of QUMOND is
\cite{milgrom:2009vn}
\begin{align}\label{eq:q}
    q=\frac{3}{2}\int_0^\infty \mathrm{d} v \int_{-1}^1 d\xi &\left(\nu -1\right)\Big[e_{\rm N}\left(3\xi -5\xi^3\right)\\ 
    &\quad +v^2\left(1-3\xi^2\right)\Big] \, , \nonumber
\end{align}
\end{subequations}
where $\nu$ in the integrand is $\nu\left[\sqrt{e_{\rm N}^2+v^4+2e_{\rm N}v^2\xi}\right]$. Computations of $q$ or $Q_2$ can be found in \cite{milgrom:2009vn,hees:2016mi,Desmond_Cassini} for the IFs considered in the present work and different values of the Newtonian external field. For the IF commonly adopted in galactic dynamics, the typical magnitude of $Q_2$ lies in the $10^{-26}$ s$^{-2}$ range.

\subsection{Impact of Jupiter}

Now, we will examine explicitly how well the hypothesis of considering only the Sun as an internal gravitational source to model the modified gravity MOND EFE in the Solar System (as done in, e.g., \cite{Milgrom2012,blanchet:2011ys,hees:2016mi,Desmond_Cassini}) is justified. For this, we consider the impact of Jupiter, the main perturbation in the Solar System, on the theoretical prediction of $Q_2$.

First, in order to visualize the problem, we use Eq.~(\ref{eq:rho_p}) to compute the phantom density distribution, which we show in Fig.~\ref{fig:phantom} for the $\delta=\gamma=1$ IF from Eq.~\eqref{eq:IF_delta} and Eq.~\eqref{eq:IF_gamma} (the RAR IF), including the contributions of both the Sun and Jupiter. In Fig.~\ref{fig:phantom}, Jupiter is chosen to be ``in Sagittarius'', and is therefore approximated as being precisely aligned along the Sun-Galactic center axis, in the direction of the Galactic center. 

As can clearly be seen on this Figure, within the Solar System ($< 100$ astronomical units -- au), the phantom density is completely negligible. This is due to the fact that the shape of the IF presents a sharp transition to the Newtonian regime at large accelerations. The phantom density becomes important only at ``MONDian'' scales, i.e. beyond several hundreds of~au. The negative phantom density zone, together with the clear non-sphericity, appears towards the MOND radius, at several thousand of~au. This non-sphericity is the source of the $Q_2$ term, see Eq.~(\ref{eq:exact}). Changing the position of Jupiter on this plot, or removing it entirely, leads to imperceptible changes on the Figure, which is a good indication that the effect of Jupiter on the phantom potential will, indeed, be completely negligible.

Let us now quantify Jupiter's effect on $Q_2$. Intuitively, since the source of the quadrupole is the asphericity of the phantom density at several thousands of~au from both the Sun and Jupiter, one can expect that it will be equivalent to merging the Sun and Jupiter at the center and simply adding up their masses. In appendix~\ref{app:Jupiter}, we compute exactly the quadrupole using Eq.~(\ref{eq:q_general}) in a situation where the position of Jupiter with respect to the Sun is aligned with the external Newtonian field. The main difference with respect to the previous calculation comes from the fact that the Newtonian field from Eq.\eqref{eq:gN} is now modified by the addition of Jupiter. The exact calculation leads to Eqs.~(\ref{eq:Q_2_jup_both}), which can be evaluated numerically. This exact result depends on two small parameters: (i) $\beta=M_\mathrm{Jup}/M_\odot\sim 10^{-3}$ and (ii) $w=D_{\odot\mathrm{-Jup}}/R_M\sim 7\times 10^{-4}$ with $D_{\odot\mathrm{-Jup}}$ the distance between the Sun and Jupiter. Expanding the exact results in terms of these two small parameters, we can then write, as intuitively expected,
\begin{align}
Q_2' \simeq
& \, Q_2 \sqrt{\frac{M_\odot}{M_\odot+M_\mathrm{Jup}}} \label{eq:delta_Q2} = Q_2 (1+ \beta)^{-1/2}\, ,
\end{align}
where $Q_2$ is the quadrupole estimate including only the Sun and $Q_2'$ is the same quantity including also the contribution from Jupiter with terms $\mathcal O(\beta w^2)=\mathcal O\left(5\times 10^{-10}\right)$ neglected. 
Taylor expanding this leads to  
\begin{equation}
    \frac{\Delta Q_2}{Q_2} \simeq -\frac{1}{2} \beta \simeq -5\times 10^{-4}\, .
\end{equation} 
In conclusion, including Jupiter's contribution always modifies the $Q_2$ estimate at a relative level of $\sim 0.05\%$, which has been confirmed using the exact formulas derived in Appendix~\ref{app:Jupiter}, i.e. using Eqs.~\eqref{eq:Q_2_jup_both}.

\subsection{Theoretical take away messages}
Considering the current observational uncertainties, on Solar System scales, the effect of AQUAL or QUMOND reduces only to a global quadrupolar contribution generated by the external field effect of the Milky Way. 

At the scales relevant for the main planets, the deviation of the MOND IF from 1 can be made completely negligible, as illustrated by the extremely small contribution of the phantom density at Solar System scales in Fig.~\ref{fig:phantom}, but the quadrupole effect remains and is entirely related to the value that the IF takes for the external field of the MW. To erase the quadrupole entirely, it is necessary to consider an IF that deviates very little from Newtonian gravity at the value of the Milky Way's external field. As we have explicitly shown here, the addition of Jupiter as a secondary internal source does not add a second quadrupole moment. All it does is producing a relative difference of the global $Q_2$ at the level of $M_\mathrm{Jup}/(2 M_\odot) \simeq 0.05\%$.

The asphericity of the Solar System also leads to an additional quadrupole contribution in AQUAL or QUMOND, even when the Solar System is considered as an isolated system : for the $\delta=\gamma=1$ IF from Eq.~\eqref{eq:IF_delta} and Eq.~\eqref{eq:IF_gamma} (the RAR IF), this effect has been shown to be at the level of $10^{-34}$ s$^{-2}$ \cite{Milgrom2012}, seven orders of magnitude below current observational limits. 

Beyond those quadrupolar contributions, the EFE also gives rise to higher multipolar moments, whose contribution relatively to the quadrupolar one has been shown to scale as $(R/R_M)^{\ell-2}$, with $\ell$ the index of the multipole and $R$ the typical size of the system considered, here 10 to 20 au for the major planets \cite{blanchet:2011ys}. For the inner Solar System dynamics, i.e. for  $(R/R_M)\leq 10^{-2}$, the $\ell=3$ octupole contribution is $\sim 100$ times  smaller than the quadrupolar one and can therefore also be safely neglected. Let us however note that, when probing the effect of MOND on, e.g., long period comets \cite{vokrouhlicky:2024aa}, one then probes more than just the quadrupole.

In conclusion, the QUMOND contribution in the \emph{inner} Solar System can safely be modeled by the modification to the Newton potential from Eq.~\eqref{eq:Phi_EFE}, which leads to an anomalous acceleration
$   \bm a_{\rm p} = Q_2\left((\hat{\bm e}\cdot \bm x)\hat{\bm e} - \bm x/3 \right)\,$. Of course, relativistic (post-Newtonian) corrections to this modeling would depend on the exact relativistic formulation of the MOND theory, but as long as the theory asymptotes to QUMOND or AQUAL at scales relevant for the Milky Way and the Solar System \cite{FamaeyDura}, they should be negligible compared to the modified Newtonian part, since they will be relatively smaller by $\mathcal O(GM_\odot/c^2R)\sim O\left(10^{-8}\right)$.

As a final theoretical remark, let us note that all exact QUMOND evaluations made here require to know the {\it Newtonian} external field from the Milky Way. It is customary \cite{hees:2016mi,Desmond_Cassini} to parametrize the predicted $Q_2$ from the value of the actual external field as measured with, e.g., Gaia \cite{gaia-collaboration:2021aa}. However, due to the asphericity of the Milky Way, the relation $a_e = \nu({a^N_e/a_0})a^N_e$ is {\it not} exact. In the following, we will consider a better approximation to this relation, from Eq.~26 of \cite{FamaeyDura} which typically modifies the $Q_2$ prediction only at the relative \% level.

\section{Solar System Data Analysis}
\label{sec:analysis}
\subsection{Data and predictions}

High-precision dynamical models of the Solar System with long-term orbital measurements provide a sensitive tool for uncovering small deviations from General Relativity expectations \cite{hees_2014,park_2017,fienga:2024aa}, such as the the modified gravity MOND quadrupole theoretically described above. The continuous refinement of planetary ephemerides, enabled by decades of radiometric tracking and optical astrometry, therefore offers an opportunity to refine existing constraints on this effect, and hence on modified gravity approaches to MOND.

Modern-day planetary ephemerides are primarily constrained by Earth-based radiometric measurements of spacecrafts. These data typically achieve a few meter-level ranging accuracy, allowing for extremely precise determinations of the orbits of both spacecraft and the bodies they orbit. Importantly, the power of such data arises not only from their instantaneous precision but also from the extended time span over which they are collected. Long-term time series, spanning years to decades, allow small, cumulative dynamical signatures to emerge from the noise \cite{park2025io,park2025small}. As a result, small deviations from Newtonian mechanics or General Relativity can be isolated. Tracking data from spacecrafts in orbit around planetary bodies are therefore particularly valuable for these investigations when ranging data are available over a substantial fraction of a planet’s orbital period (e.g. typically more than a quarter). The resulting orbit determination then achieves exceptional recovery.

As explained in depth in the previous Section, the relevant dynamical effect in modified gravity MOND at the scale of planetary orbits can be parameterized by the quadrupole term, $Q_2$, see Eq.~\eqref{eq:Phi_EFE}. Previously, this parameter was constrained using Cassini radio tracking data collected between May 2004 and April 2013 \cite{hees_2014}. In the present study, we revisit and extend that analysis. The Cassini data arc is augmented by an additional three years of observations, now extending through August 2017, enhancing the temporal coverage and improve sensitivity to small secular trends \citep[DE440, see][]{park_2021}.

\begin{figure}[htb]
  \centering
  \includegraphics[width=0.5\textwidth]{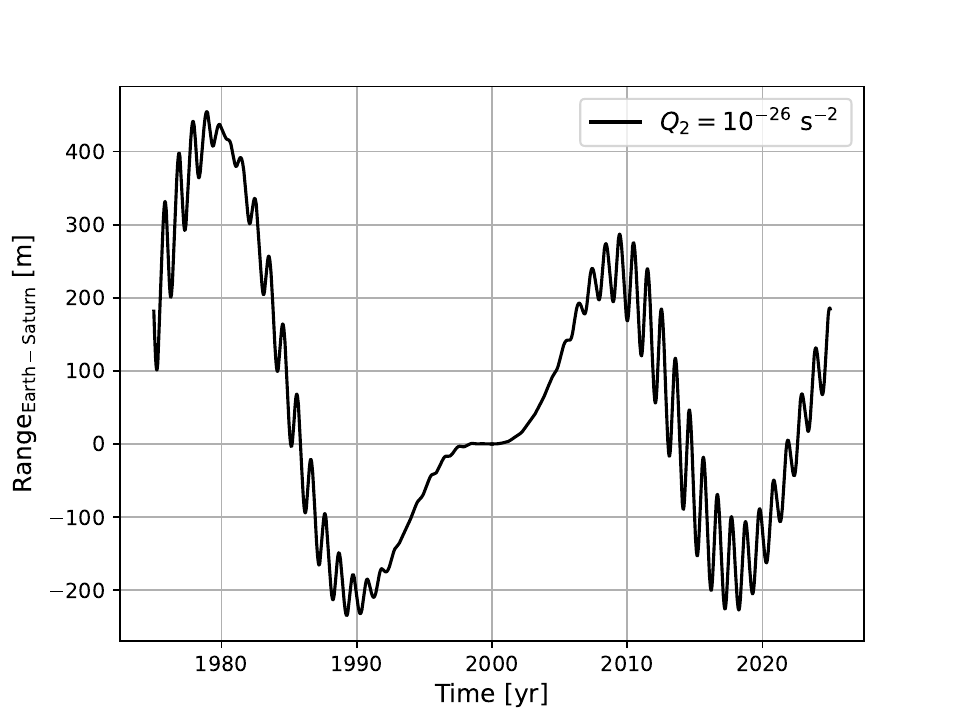}
  \caption{Simulation of the deviation of the Earth-Saturn 1-way range over time when including a quadrupole as in Eq.~\eqref{eq:Phi_EFE} with $Q_2 = 10^{-26} {\rm s}^{-2}$.}
  \label{fig:range}
\end{figure}

Before observationally estimating the MOND $Q_2$ parameter by simultaneously fitting it with all other parameters included in the planetary ephemerides, we give in Fig.~\ref{fig:range} an estimate of the deviation over time of the Earth-Saturn 1-way range when including a quadrupole potential as in Eq.~\eqref{eq:Phi_EFE} with $Q_2 = 10^{-26} {\rm s}^{-2}$, which is the typical order of magnitude predicted from MOND IFs that fit galaxy rotation curves well \cite{Desmond_Cassini}. The typical deviation is of the order of several hundreds of meters with oscillations at both the Earth and Saturn orbital frequencies.

\subsection{Modeling and results}\label{sec:modelin_results}

\begin{figure*}[htb]
  \centering
  \includegraphics[width=1.0\textwidth]{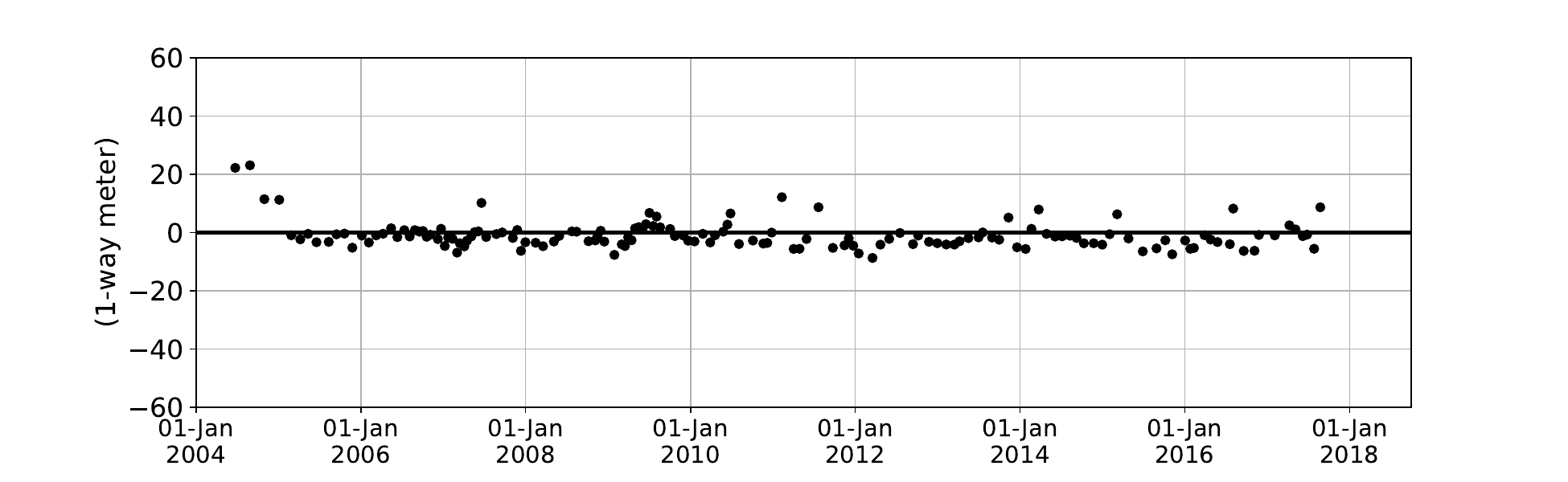}
  \caption{Residuals of the Cassini range data against the ``All Data" case. The weighted root-mean-square residual is about 4 m.}
  \label{fig:cassini_resid}
\end{figure*}

Our analysis follows procedures analogous to those used in constructing JPL’s Development Ephemerides (DE) series \cite{folkner_2014,park_2021}. The DE framework employs a parameterized post-Newtonian (PPN) formulation for point-mass interactions, including relativistic corrections and various gravitational perturbations. This study includes all Cassini data plus all-available data used to compute the latest DE series, i.e., DE440 (see \cite{folkner_2014,park_2021} for more details).  Similar to prior studies \cite{hees_2014}, we incorporated the quadrupole term directly into the equations of motion and estimated the $Q_2$ parameter in a comprehensive, global least-squares fit that \emph{simultaneously} solved for all the standard DE parameters, totaling 792 parameters in the  estimation setup.

Table \ref{tab:table1} shows the recovered $Q_2$ values from three different fits. The “All Data” case corresponds to the $Q_2$ value recovered based on all available data used to compute DE440 \cite{ park_2021}, including all Cassini ranging data. In order to validate our results, we then also estimate this parameter using two fully independent subsets of the Cassini data to check whether the confidence interval of the estimate is stable. The “Subset 1” case is based on all available data plus the Cassini ranging data obtained between May 2004 and June 2012. The “Subset 2” case is based on all available data plus the Cassini ranging data obtained between July 2012 and August 2017. Note that the ``Subset" cases in this study and in \cite{hees_2014} use different Cassini ranging data arcs. The data arc considered in our “Subset 1” case is actually close but different to the ``All data" case in \cite{hees_2014} (i.e., $\sim$8 years vs. $\sim$9 years). Both ``Subset 1" and ``Subset 2" cases in our study also use all available data used to compute DE440 together with the Cassini post-reconstruction orbit SAT441 \cite{jacobson2022orbits}. We focus on these subsets of the Cassini data because they provide the most accurate ranging measurements to the most distant planet in the Solar System with long-term radiometric tracking in the data set, where the quadrupole MOND effect would be expected to be most pronounced. 

Our recommended solution is the ``All Data" case $Q_2=(1.6\pm1.8)\times 10^{-27}$ s$^{-2}$$~(1\sigma)$, which is consistent with zero within the measurement uncertainty and represents a 40\% improvement in accuracy relative to the previous study result of $Q_2=(3\pm3)\times 10^{-27}$ s$^{-2}$~$(1\sigma)$ \cite{hees_2014}. The variations of the solutions shown in Table \ref{tab:table1} are fully consistent with statistical fluctuations expected from measurement noise and parameter correlations, indicating the stability and robustness of the recovered $Q_2$ estimate. The uncertainties are smaller than those reported in the previous study \cite{hees_2014}, primarily because longer Cassini data arcs are considered combined with the complete data set used to compute DE440. Figure \ref{fig:cassini_resid} shows the residuals of the Cassini range data relative to this ``All Data" case, indicating a fit consistent with DE440 at the few meter level. We note that a large part of the signature of a non-zero $Q_2$ has been reabsorbed in the fit due to correlations with other estimated parameters, see \cite{Hees:2012} for a detailed discussion. However, including the $Q_2$ parameter as an additional parameter in the fit did not improve the Saturn residuals, compared to Fig.~11 of \cite{park_2021}. 

The persistence of this null result for $Q_2$ across independent data sets and extended time spans further strengthens the empirical limits on modified gravity MOND-like deviations within the Solar System and underscores the continued value of long-term tracking of planetary motions as a probe of fundamental physics.

\begin{table}[htb]
\caption{\label{tab:table1}%
Estimates of $Q_2$ and related uncertainties (1-$\sigma$) based on all and subsets of Cassini radio tracking data (see Sec.~\ref{sec:modelin_results} III B for a description of the subsets).
}
\begin{ruledtabular}
\begin{tabular}{lcc}
\textrm{Cases}&
$Q_{2}[10^{-27}\textrm{s}^{-2}]$&
$\sigma_{Q_{2}}[10^{-27}\textrm{s}^{-2}]$\\
\colrule
$\textrm{All Data}$ & 1.6 & 1.8 \\
$\textrm{Subset 1 of Cassini}$ &  -1.6 & 3.6 \\
$\textrm{Subset 2 of Cassini}$ & -2.4 & 3.9 \\
\end{tabular}
\end{ruledtabular}
\end{table}

\section{Updated constraints on MOND as modified gravity} \label{sec:results}

\subsection{Constraints from adjusting the IF and acceleration scale to SPARC data}\label{sec:SPARC}

We now update the constraints of~\cite{Desmond_Cassini} obtained using the SPARC galaxy rotation curves dataset \cite{lelli:2016} with the new $Q_2$ bound from the previous section. This work is built on \cite{uRAR}, which infers properties of the MOND IF while marginalising over all relevant galaxy-specific nuisance parameters, including distances, inclinations, luminosities and disk, bulge and gas mass-to-light ratios. As in~\cite{Desmond_Cassini} we investigate the three IF families from Eqs.~(\ref{eq:if_fams}), as well as seven priors and hyperpriors for the mass-to-light ($M/L$) ratios of the SPARC \cite{SPARC} rotation curve data and models in order to compare the constraints on MOND parameters between external galaxy rotation curves and the Cassini quadrupole. The seven $M/L$ models are identical to the ones considered in \cite{Desmond_Cassini} and include the fiducial lognormal priors from SPARC (``Fiducial''), allowing the centers of these fiducial $M/L$ priors to float as hyperpriors (``Free hyper''), requiring the center of the disk $M/L$ to be smaller than that of the bulge (``Constrained hyper''), $M/L$ drawn independently from wide uniform priors (``Free unif''), the same as the latter but requiring a larger bulge $M/L$ galaxy-by-galaxy (``Constrained unif''), free hyperpriors but removing the 31 galaxies with bulges (``No bulge''), free hyperpriors but removing the 116 pure disk galaxies without bulges (``Only bulge''). 

We consider two cases without any EFE, one in which the radial acceleration in the disk is taken to be $a_R = \nu({a^N_R/a_0})a^N_R$, which cannot be exact within a disk in modified gravity MOND (but can be, for circular orbits, in modified inertia) and another one where a better approximation is taken to be \cite{Brada:1995,FamaeyDura}
\begin{subequations}\label{eq:eN_Sigma}
\begin{equation}\label{eq:nu_eN}
    \left|\bm a_R\right| = \nu\left(\frac{a^\dagger_{NR}}{a_0}\right) \left|\bm a_{R}^N\right|\, , 
\end{equation}
with 
\begin{equation}
    a^\dagger_{NR} = \left(\left|\bm a_{R}^N\right|^2 + \left(2\pi G \Sigma_b\right)^2\right)^{1/2}\, ,
\end{equation}
\end{subequations}
with $\Sigma_b$ the local baryonic disk surface density. We test this case with the fiducial $M/L$ model. In addition, we also test the impact of the EFE on galaxy rotation curves using two prescriptions from \cite{chae:2022tt}: the ``Freundlich--Oria analytic'' model~\citep{Freundlich,Oria} for QUMOND and the ``AQUAL numerical'' model, both of which act as fitting functions for RCs by modifying the raw IFs and can therefore be applied to any IF. Writing the Newtonian external field strength in units of $a_0$ as $e_{\rm N} \equiv a^N_e/a_0$, the QUMOND EFE model is
\begin{equation}\label{eq:efe_fo}
\nu_\text{EFE, QUMOND}(y)=\nu\!\left({\rm min}\!\left[y+\frac{e_{\rm N}^2}{3y},\,e_{\rm N}+\frac{y^2}{3e_{\rm N}}\right]\right),
\end{equation}
while the AQUAL model is
\begin{equation}\label{eq:efe_an}
\nu_\text{EFE, AQUAL}(y)=\nu(y_\zeta)\!\left[1+\tanh\!\left(\frac{\zeta e_{\rm N}}{y}\right)^{\chi}\frac{\hat{\nu}(y_\zeta)}{3}\right],
\end{equation}
with $\hat{\nu}(y)\equiv{\rm d}\ln\nu/{\rm d}\ln y$ and $y_\zeta\equiv\sqrt{y^2+(\zeta e_{\rm N})^2}$. We take the best-fit values $\zeta=1.1$ and $\chi=1.2$ from \cite{chae:2022tt}. For each EFE prescription we consider two treatments of $e_{\rm N}$: a single global value inferred from the data with a wide uniform prior, and a galaxy-by-galaxy local value with a prior given by the ``maximum clustering'' model of \cite{Chae_2}, which estimates $e_{\rm N}$ from the large-scale baryonic environment under the assumption that unseen baryons correlate maximally with observed ones; each galaxy’s $e_{\rm N}$ is then marginalised over. For further details see~\cite{Desmond_Cassini}, to which our present modeling here is identical.

\begin{figure*}
  \centering
  \includegraphics[width=0.32\textwidth,height=0.33\textwidth]{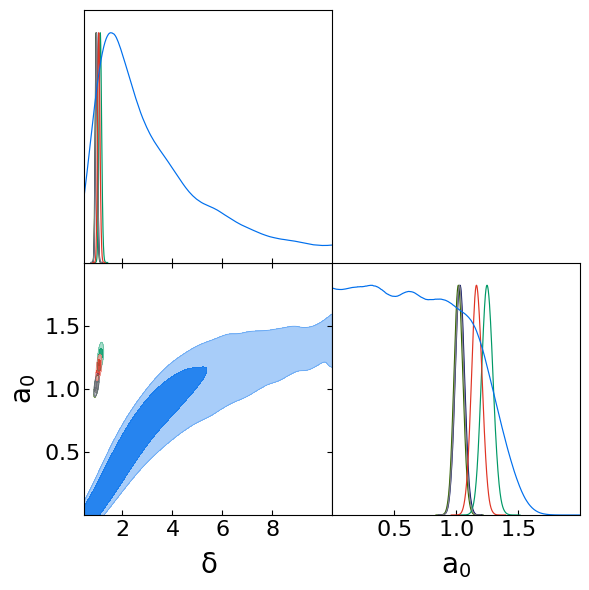}
  \includegraphics[width=0.32\textwidth,height=0.33\textwidth]{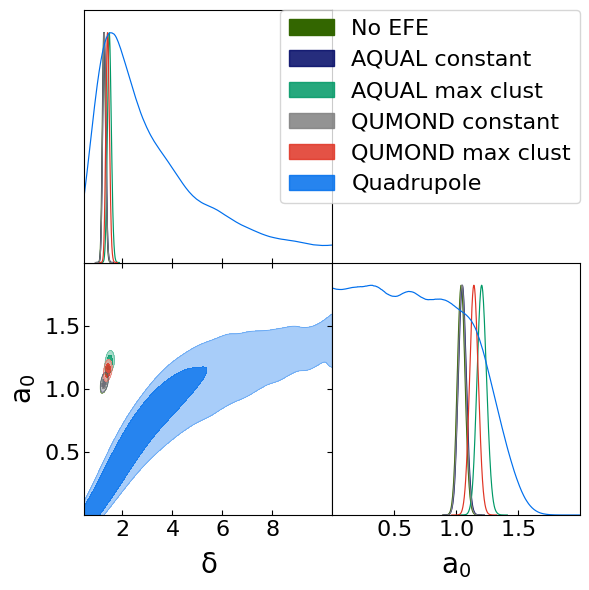}
  \includegraphics[width=0.32\textwidth,height=0.33\textwidth]{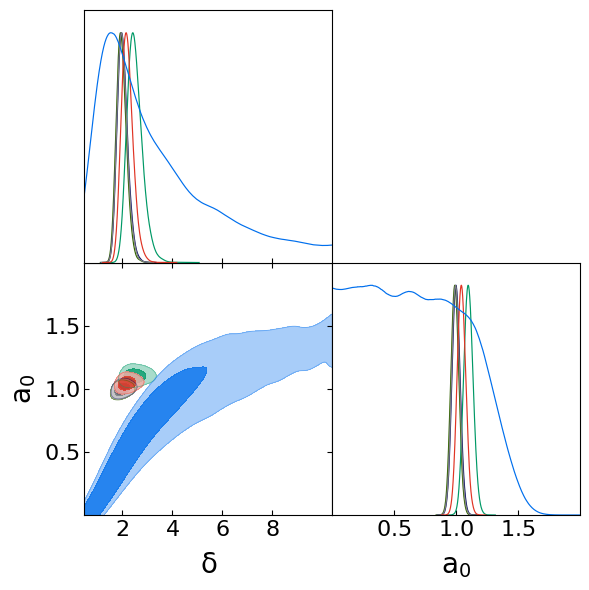}
  \caption{Posteriors on $a_0$ (in $10^{-10}$ m s$^{-2}$) and $\delta$ from the RAR and SS quadrupole. \emph{Left:} Fiducial SPARC $M/L$ model; \emph{center:} free Gaussian hyperprior model; \emph{right:} as center but excluding galaxies with bulges.}
  \label{fig:getdist}
\end{figure*}

For the $\delta$-family, Fig.~\ref{fig:getdist} compares the constraints on $a_0$ and $\delta$ from the updated $Q_2$ measurement (blue) and the RAR (other colours). The different panels show different SPARC $M/L$ models while the different RAR contours within each panel vary the EFE model (the Cassini constraint is the same in each case). We focus on $M/L$ models 1.a., 2. and 6. in Table~\ref{tab:results}. Compared to the analogous Fig.~9 in~\cite{Desmond_Cassini} the blue contour is somewhat flatter, reflecting the larger $\delta$ (i.e. sharper MOND--Newton transition) required at given $a_0$ to accommodate the strengthened $Q_2$ measurement. While some of the EFE models overlapped at the $2\sigma$ level with the Cassini contour in the right-hand panel in~\cite{Desmond_Cassini}, this is no longer the case.

Table~\ref{tab:results} shows the level of tension between the RAR and $Q_2$ constraints; following Tables 1 and 2 of~\cite{Desmond_Cassini} we focus on the no-EFE and galaxy-by-galaxy AQUAL models. The sigma level is determined by calculating the posterior predictive distribution of $Q_2$ from each set of RAR posteriors and then assuming Gaussianity to sum the RAR and Cassini $Q_2$ uncertainties in quadrature. We see enhanced tensions relative to~\cite{Desmond_Cassini}, roughly $9\sigma \rightarrow 15\sigma$ for the fiducial SPARC $M/L$ model and $7\sigma \rightarrow 12\sigma$ when freeing up the $M/L$ prior centres. The no-bulge models that gave consistency before (2.7 or 1.9$\sigma$ depending on the EFE) now give moderate tension (3.8$\sigma$ or 2.8$\sigma$).

We note that SPARC is not the only galaxy rotation curves dataset constraining the RAR, although the only other RAR measurement of $a_0$ and $\delta$ (allowing for comparison with the Cassini measurement) is from the MIGHTEE survey \cite{Varasteanu}. Here a higher value of $\delta$ is inferred, which would generate less tension with the $Q_2$ constraint. We leave a more detailed comparison to future work.

\begin{table}[htb]
\begin{center}
\begin{tabular}{|c|c|c|}
  \hline
  $M/L$ model & $\sigma_{Q_2}$ (no EFE) & $\sigma_{Q_2}$ (AQUAL local) \\
  \hline
  \rule{0pt}{4ex}
  1.a. Fiducial & 15.0 & 15.2\\
  \rule{0pt}{4ex}
    1.b.  w. surf. density & 15.8 & -\\
  \rule{0pt}{4ex}
  2. Free hyper & 12.0 & 11.9\\
  \rule{0pt}{4ex}
  3. Constrained hyper & 12.6 & 12.5\\
  \rule{0pt}{4ex}
  4. Free unif & 12.6 & 12.1\\
  \rule{0pt}{4ex}
  5. Constrained unif & 15.3 & 15.5\\
  \rule{0pt}{4ex}
    6.  No bulge & 3.8 & 2.8\\
  \rule{0pt}{4ex}
  7. Only bulge & 15.9 & 15.9\\
  \hline
\end{tabular}
\caption{The level of tension between various RAR models and the updated Cassini $Q_2$ constraint. The 7 models considered are described in Sec.~\ref{sec:SPARC}.}
\label{tab:results}
\end{center}
\end{table}

\begin{figure}
  \centering
  \includegraphics[width=0.45\textwidth]{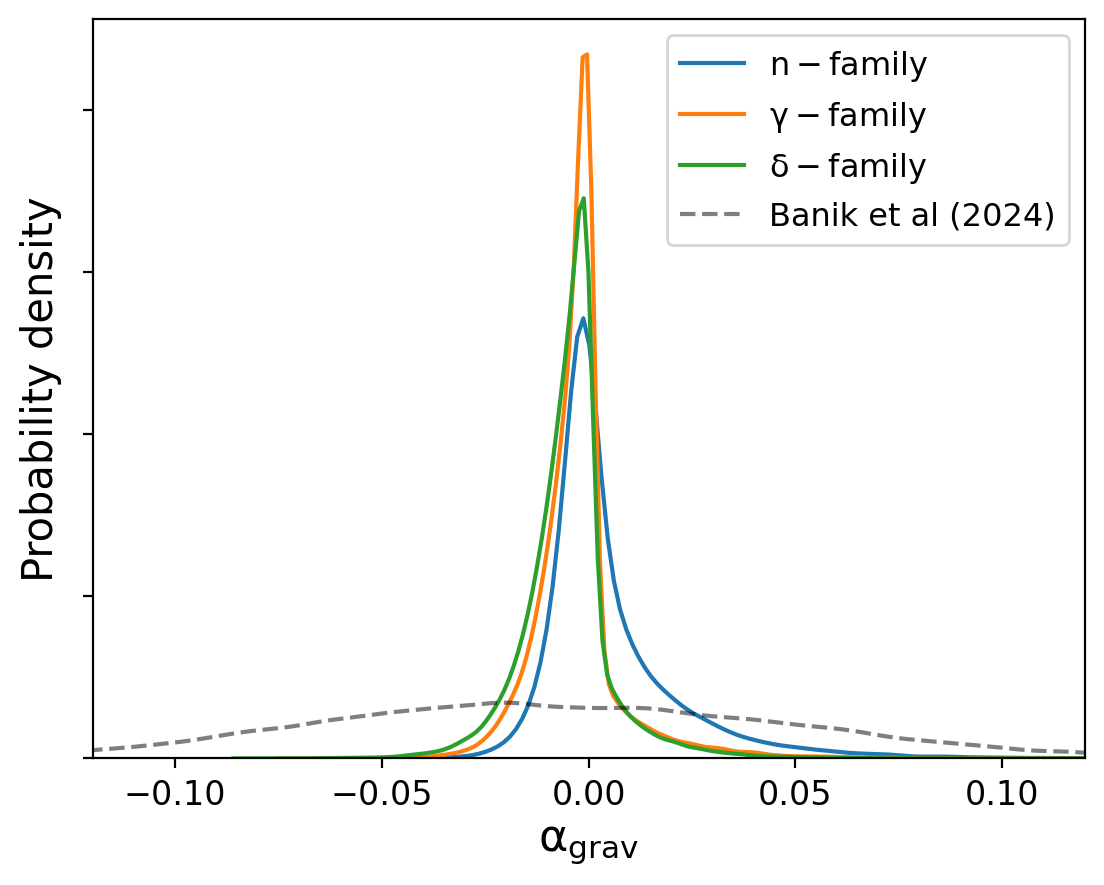}
  \caption{Posterior predictive distribution of $\ag$ for the three IF families from the updated $Q_2$ constraint.}
  \label{fig:agrav_q2}
\end{figure}

\subsection{The local $Q_2$ constraint and the Milky Way}
The $Q_2$ constraint being based on the Solar System itself being embedded in the Milky Way galaxy, it is in principle possible to consider \emph{only} the gravitational field (or rotation curve) constraints of the Milky Way itself, instead of external galaxies such as the SPARC dataset, in order to self-consistently test QUMOND predictions in our Galaxy. 

Therefore, we now explore the connection between the local $Q_2$ constraint and other measurements performed in the Milky Way. First of all, we show that starting from the Gaia \cite{gaia-collaboration:2021aa} direct measurement of the acceleration at the Sun, and converting it into a constraint on the Newtonian external field by including the surface density term of Eqs.~\eqref{eq:eN_Sigma} (which is a better approximation than the customary spherical approximation) only impacts the $Q_2$ estimate by less than  1\%. Second, we update our discussion of \cite{Desmond_Cassini} on the tension with respect to results from Gaia wide binaries claiming to detect a QUMOND or AQUAL signature. Finally, we present a self-consistent test of MOND as modified gravity within the Milky Way, by re-interpreting the local $Q_2$ constraint as a constraint on the local ratio $a_e/a^N_e$, which can be re-casted into a strong constraint on the mass modeling of the Milky Way rotation curve at the position of the Sun.

\subsubsection{Impact of the Milky Way's non-sphericity}
The QUMOND quadrupole parameter computed in this paper depends on the {\it Newtonian} gravitational field $\bm e_N = \bm a_e^N/a_0$ which appears in the integrand of Eq.~\eqref{eq:q}. This Newtonian field is most often \cite{hees:2016mi,Desmond_Cassini} itself estimated from actual measurements of the external gravitational field $\bm a_e$ \cite{gaia-collaboration:2021aa} instead of less well constrained direct measures of the baryonic radial acceleration (that depend on the exact distribution of baryons, stellar $M/L$ ratio, etc). The conversion is usually made by using the relation $a_e = \nu({a^N_e/a_0})a^N_e$, which is \emph{not} exact for a flattened galactic disk in QUMOND.

Another better approximation (although still not exact) is given \cite{FamaeyDura} by Eqs.~\eqref{eq:eN_Sigma}. We can therefore estimate the impact of the Milky Way baryonic surface density term for the above conversion $a_e \rightarrow a^N_e$, and the related $Q_2$ prediction in QUMOND. The external acceleration of the Sun is $a_e = 2.32\times 10^{-10}$ m s$^{-2}$ \cite{gaia-collaboration:2021aa} whilst the best-fit MOND acceleration scale for the $\nu_{\delta=1}=\nu_{\gamma=1}$ IF is $a_0=1.02\times 10^{-10}$  m s$^{-2}$ (see Tab. 1 from \cite{Desmond_Cassini}). Taking $\Sigma_b=47 \, {\rm M}_\odot$/pc$^2$ \cite{bland-hawthorn:2016aa} in Eqs.~\eqref{eq:eN_Sigma} gives a Newtonian field $e_N=1.659$, slightly larger than its counterpart computed under the assumption of spherical symmetry ($e_N=1.643$). This leads to an increase of the $Q_2$ parameter by less than $1\%$. More precisely, $Q_2=3.387\times 10^{-26}$ s$^{-2}$ when neglecting the surface density and is $Q_2=3.411\times 10^{-26}$ s$^{-2}$ when accounting for the surface density.

\subsubsection{Consistency with Gaia wide binaries}
We now also revisit the prediction of the outcome of the wide binary test from the $Q_2$ posterior, as done in \cite{Desmond_Cassini}. This is parametrized by
\begin{equation}\label{eq:ag}
\ag \equiv \frac{\sqrt{\eta}-1}{0.193},
\end{equation}
where $\eta \equiv \langle a_R \rangle / a^N_R$, the ratio of the 
azimuthally averaged asymptotic radial gravity to the Newtonian expectation~\citep{Banik_WBT}. This is chosen such that $\ag=0$ for Newtonian gravity, and $\ag=1$ for the case of QUMOND with the $n=1$ IF assuming $a_0=1.2\times10^{-10}$ m s$^{-2}$ and $a_e= 1.8a_0$, explaining the numerical value in the denominator of the definition of $\alpha_\textrm{grav}$.

The posterior predictive distributions of $\ag$ for the $n-$, $\delta-$ and $\gamma-$family IF fits to the $Q_2$ constraint are shown in Fig.~\ref{fig:agrav_q2}. The constraint from the wide binary test of~\cite{Banik_WBT} is shown in dashed gray. Compared to the analogous Fig.~10 of~\cite{Desmond_Cassini}, we see a considerably tightening of the $Q_2$-based posteriors around $\ag=0$, consolidating the fact that the Cassini constraint implies (in the context of QUMOND and AQUAL) a null test for $\ag$ to much higher precision than the Gaia wide binary data itself affords. This does not {\it per se} mean that there is no deviation from Newtonian gravity in the dynamics of wide binaries. But it means that such a deviation cannot be considered to be an effect of QUMOND or AQUAL once the Solar System constraints are taken into account. 

\subsubsection{A self-consistent test within our Galaxy}\label{sec:test_MOND_MW}
As already discussed in depth in \cite{Desmond_Cassini}, the $Q_2$ constraint actually probes the behavior of the IF at accelerations of the order of the acceleration of the Solar System within the Galaxy. Since the value of the IF, once taking into account a surface density term as in Eqs.~\eqref{eq:eN_Sigma}, is almost exactly a direct measure of $a_e/a^N_e$, the $Q_2$ constraint can be reinterpreted as a constraint on $a_e/a^N_e$, hence on the MOND boost to the radial acceleration at the position of the Sun in the Milky Way.

This is very similar to computing the posterior on $\ag$, although $\eta$ for wide binaries is not exactly the same as $a_e/a^N_e$ \citep[see, e.g., eq.14 of][]{Desmond_Cassini}. So we can intuitively expect that the $Q_2$ constraint will imply a very small value of $a_e/a^N_e -1 $, which we will now quantify as self-consistently as possible, by also including the surface density term. The $Q_2$ parameter depends on 4 parameters, see Eqs.~\eqref{eq:Q2_sun_both} and~\eqref{eq:eN_Sigma}: (i) the IF parametrized by a shape parameter, see Eqs.~\eqref{eq:if_fams}, (ii) the MOND acceleration scale $a_0$, (iii) the acceleration field of the Milky Way at the position of the Sun, $a_e$, and (iv) the value of the baryonic surface density $\Sigma$. 

\begin{figure}
  \centering
  \includegraphics[width=0.40\textwidth]{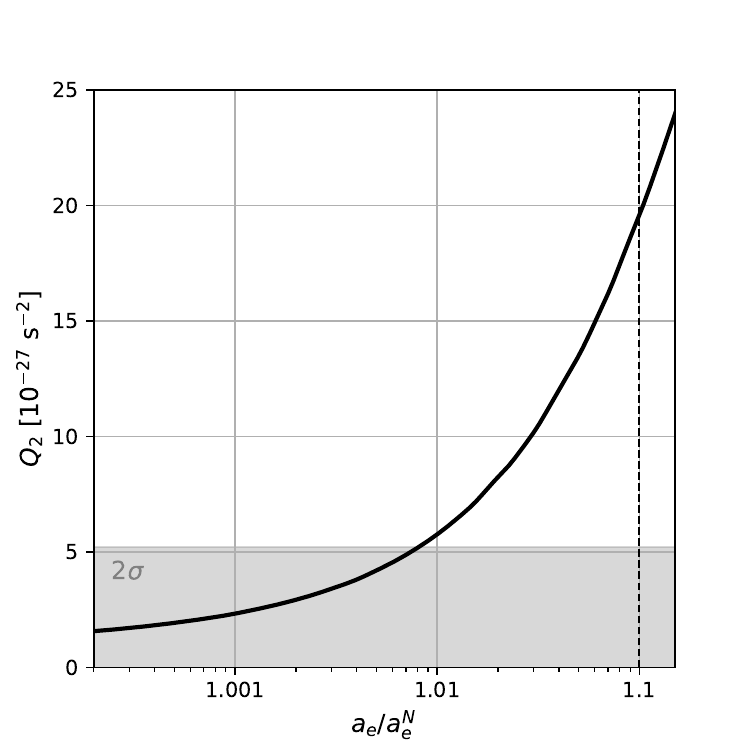}
  \caption{Evolution of the $Q_2$ parameter for the $\delta$-family IF from Eq.~\eqref{eq:IF_delta} as a function of $a_e/a^N_e$. The $a_0$, $a_e$ and $\Sigma$ parameters used to estimate $Q_2$ are fixed to fiducial values (see text). Each point corresponds to a different value of $\delta$. The shaded area corresponds to the 2$\sigma$ constraint from Sec.~\ref{sec:analysis} and the vertical dashed line to a conservatively low observational estimate of $a_e/a_e^N$ \cite{Sofue:2025}.}
  \label{fig:MW}
\end{figure}

First, for illustrative purposes, Fig.~\ref{fig:MW} shows the value of the $Q_2$ prediction as a function of $a_e/a^N_e$ for the $\delta$-family IF, using fixed fiducial parameters $a_0 = 1.02\times 10^{-10}$ m s$^{-2}$ \cite{Desmond_Cassini}, $a_e = 2.32\times 10^{-10}$ m s$^{-2}$ \cite{gaia-collaboration:2021aa} and $\Sigma = 47 \, M_\odot$/pc$^2$ \cite{bland-hawthorn:2016aa}. Each point of this curve corresponds to a different value for the IF shape parameter $\delta$, which translates into a value of $a_e/a^N_e$ through Eqs.~\eqref{eq:eN_Sigma}. The gray area corresponds to the 2$\sigma$ $Q_2$ constraint from Sec.~\ref{sec:analysis}. The vertical dashed line corresponds to a conservatively low observational estimate on $a_e/a_e^N$ from the Milky Way rotation curve \cite{Sofue:2025}. This figure clearly illustrates that for a given family of IF, the local $Q_2$ constraint translates into a constraint on the ratio of the galactic parameters $a_e/a^N_e$. Furthermore, it also illustrates that the $Q_2$ constraint is in strong tension with galactic measurements of $a_e/a_e^N$ ($\sim 11\sigma$ tension). Only a percent-level boost is allowed from the $Q_2$ constraint, but it is measured to be at least at the 10\% level from the Milky Way rotation curve \cite{Sofue:2025}.

\begin{figure}
  \centering
  \includegraphics[width=0.4\textwidth]{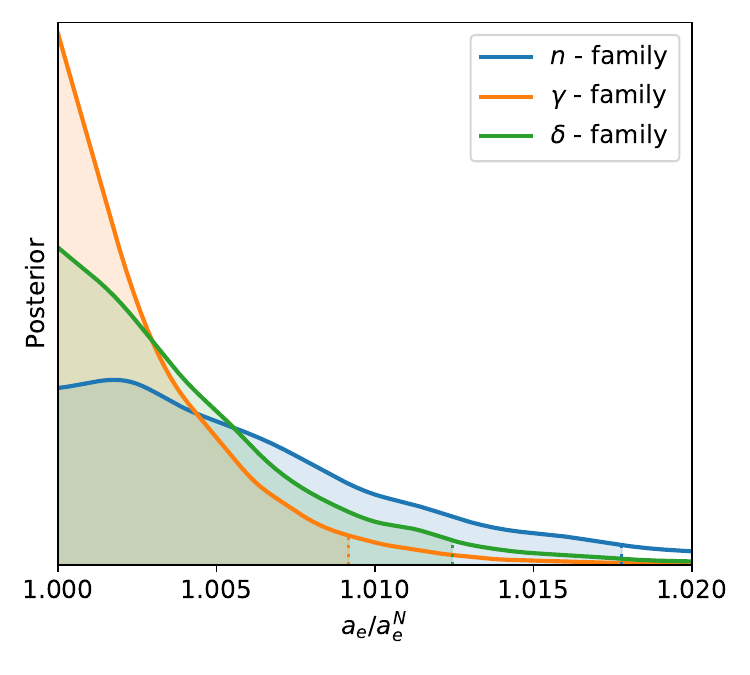}
  \caption{Posterior probability distribution of $a_e/a^N_e$ obtained from the local $Q_2$ constraint and marginalized over $a_0$, $a_e$ and $\Sigma$ for the three families of IF considered in this work. The shaded areas correspond to the 95\% credible intervals.}
  \label{fig:post_MW}
\end{figure}

To obtain a full posterior on $a_e/a^N_e$ from the $Q_2$ constraint, we performed a Bayesian inference fitting on the 4 parameters $a_0$, $a_e$, $\Sigma$ and $a_e/a^N_e$ using a Gaussian likelihood. We use a flat prior on $a_0$ between $0.5$ and $2\times10^{-10}$ m s$^{-2}$, a Gaussian prior on $a_e$ based on the Gaia estimate, $a_e = (2.32\pm0.16) \times 10^{-10}$ m s$^{-2}$ \cite{gaia-collaboration:2021aa}, a Gaussian prior on the surface density, $\Sigma= 47\pm 3 M_\odot/$pc$^2$ \cite{bland-hawthorn:2016aa}, and a wide flat prior on $a_e/a^N_e$. The resulting posterior probability distributions for the three IF families considered in this work is presented in Fig.~\ref{fig:post_MW}. The shaded areas represent the 95\% credible intervals, which lead to upper limit on $a_e/a^N_e$ of 1.018 for the $n$-family, of 1.01 for the $\gamma$-family and of 1.012 for the $\delta$-family. We checked that these posteriors are not significantly sensitive to the particular choice of prior for $a_0$ and $a_e/a^N_e$. In particular, using a flat prior on $Q_2$ similarly to \cite{Desmond_Cassini} instead of a flat prior on $a_e/a^N_e$ reduces the 95\% upper limits at a relative level of $10^{-2}$.

This result means that for the three families of IF considered in this work, the local $Q_2$ constraint obtained in Sec.~\ref{sec:analysis} implies a radial MOND boost (i.e. the multiplicative boost to the Newtonian acceleration) not larger than $\sim 1.02$ at the location of the Solar System in the Milky Way. This has been estimated observationally to $a_e/a^N_e\approx 1.1$ \cite{Sofue:2025}, or even above in most other rotation curve decompositions \citep[see, e.g.,][]{McGaugh2015}, in strong ($> 10\sigma$) tension with the posteriors from Fig.~\ref{fig:post_MW}.

\section{Conclusions}
We presented an updated observational constraint on the Solar System quadrupole parameter $Q_2$ from Eq.~\eqref{eq:Phi_EFE}, which quantifies the external field effect predicted by modified gravity MOND in the inner Solar System \cite{milgrom:2009vn}. Using the full dataset employed to compute DE440 planetary ephemerides \cite{park_2021}, including an extended Cassini dataset through 2017, we found $Q_2 = (1.6 \pm 1.8) \times 10^{-27}$~s$^{-2}$, improving the previous estimate by 40\%. An accurate theoretical modeling demonstrates that planetary contributions are negligible in predicting the MOND $Q_2$, with Jupiter accounting for only a 0.05\% effect, validating the approximation of a Sun-only internal gravitational source. Approximating the Newtonian baryonic galactic acceleration field from the measured total acceleration field, by including or not a surface density term to account for the flattening of the disk, also impacts the $Q_2$ prediction by only $\sim 1$ \%. 

Comparison of this Solar System constraint with SPARC galactic rotation curve data \cite{lelli:2016} confirms significant tension \cite{Desmond_Cassini}, as most MOND IFs that reproduce rotation curves well predict $Q_2$ values much higher than the Solar System limit. Furthermore, this $Q_2$ constraint, although obtained using local Solar System measurements provides a self-consistent test of modified gravity MOND in the Milky Way. Indeed, as explicitly shown in \cite{Desmond_Cassini}, the $Q_2$ constraint probes the IF at acceleration scales corresponding to the external galactic gravitational field. We showed that the $Q_2$ constraint can be re-interpreted as a constraint on the MOND boost (ratio of the observed gravitational acceleration to the baryonic Newtonian acceleration at the location of the Solar System) for a given IF family leading to an upper limit on this MOND boost at the level of 1.01-1.02 (95\% confidence) for the three IF-families considered in this work, see Eqs.~\eqref{eq:if_fams}. This constraint is in strong tension with the Milky Way rotation curve, from which the boost is at a level of 1.1 or more. Correlatively, the updated $Q_2$ posterior also predicts a negligible effect in wide binary systems when interpreted in the context of AQUAL or QUMOND, confirming that Solar System measurements currently provide the most stringent tests of MOND-like deviations from Newtonian gravity. This is in agreement with \cite{Banik_WBT} and in disagreement with \cite{Hernandez,*Hernandez_2,*Hernandez_4,*Hernandez_Chae}. This does not mean, {\it per se}, that there is no deviation from Newton in the dynamics of wide binaries, but it means that such a deviation cannot be considered to be a prediction of QUMOND or AQUAL, once the Solar System constraints are taken into account. 

As a conclusion, the discrepancies between galactic measurements -- especially in the Milky Way -- and the Solar System $Q_2$ constraint obtained in Sec.~\ref{sec:analysis} challenge MOND modified gravity models which boil down to the AQUAL or QUMOND formulation in the Newtonian limit. This implies that any MOND-based modified gravity must involve another scale beyond $a_0$, such a length or frequency \cite{GQUMOND}, or a screening mechanism, such as the Vainstein mechanism \cite{babichev:2011uq}. Another viable possibility is for MOND to be a modification of inertia rather than gravity \cite{inertia,inertia2023}.

\section*{Acknowledgments}
This research was carried out in part at the Jet Propulsion Laboratory, California Institute of Technology, under a contract with the National Aeronautics and Space Administration. HD is supported by a Royal Society University Research Fellowship (grant no. 211046).

\bibliography{refs}
\appendix
\section{Impact of Jupiter on the quadrupole}\label{app:Jupiter}
In this appendix, we will quantify Jupiter's effect on $Q_2$ following the same calculation as the one sketched in Sec.~\ref{sec:Q2_solar_system} using Eq.~(\ref{eq:q_general}). The main difference with respect to the calculation considering only the Sun comes from the fact that the Newtonian field is now given by
\begin{equation}\label{eq:gN_Jup}
    \frac{\bm a^N}{a_0} =  e_N \hat{\bm e} - \frac{GM_\odot}{\left|\bm x-\bm x_\odot\right|^3}(\bm x-\bm x_\odot)  -  \frac{GM_\mathrm{Jup}}{\left|\bm x-\bm x_\mathrm{Jup}\right|^3} (\bm x-\bm x_\mathrm{Jup}) \, ,
\end{equation}
with $\bm x_\odot$ and $\bm x_\mathrm{Jup}$ the positions of the Sun and Jupiter, in a coordinate system whose origin coincides with the barycenter, such that the dipolar term again vanishes exactly. As in Fig.~\ref{fig:phantom}, let us consider the case where Jupiter's position would be aligned with the direction of the external Galactic field. The MOND radius now depends on Jupiter's mass,
$R_M' = \sqrt{G(M_\odot +M_\mathrm{Jup})/a_0}$,
where the prime denotes a quantity evaluated when including Jupiter. We also introduce rescaled positions
\begin{equation}
    \bm u = \frac{\bm x}{R'_M} \, , \qquad \mathrm{and}\qquad \bm x_\mathrm{Jup} -\bm x_\odot = w R_M'\hat {\bm e} \, ,
\end{equation}
where $w$ is the Jupiter-Sun distance expressed in unit of $R_M'$ which is $\sim 6.8\times 10^{-4}$. Using these notations, the Newtonian field can be written as
\begin{equation}\label{eq:gN_Jup2}
    \frac{\bm a^N}{a_0} = e_N' \hat{\bm e} - \alpha' \frac{\bm u}{\left|\bm u\right|^3}   \, ,
\end{equation}
where 
\begin{widetext}
\begin{subequations}\label{eq:tilde_e_alpha}
\begin{align}
   e_N' &= e_N - \frac{w \beta}{(1+\beta)^2} \left[\frac{1}{\left|\bm u +\frac{\beta}{1+\beta} w\hat{\bm e} \right|^3} - \frac{1}{\left|\bm u - \frac{1}{1+\beta} w\hat{\bm e} \right|^3}\right]  \nonumber \\
  & = e_N - \frac{w \beta v^3}{(1+\beta)^2} \left[\frac{1}{\left(1+ 2 \frac{\beta}{1+\beta}vw\xi + \left(\frac{\beta}{1+\beta}\right)^2w^2v^2\right)^{3/2}} -\frac{1}{\left(1- 2 \frac{1}{1+\beta}vw\xi + \left(\frac{1}{1+\beta}\right)^2w^2v^2\right)^{3/2}}\right] \, , \\
   \alpha' & = \frac{1}{1+\beta}\frac{\left|\bm u\right|^3}{\left|\bm u +\frac{\beta}{1+\beta} w\hat{\bm e} \right|^3} + \frac{\beta}{1+\beta} \frac{\left|\bm u\right|^3}{\left|\bm u - \frac{1}{1+\beta} w\hat{\bm e} \right|^3} \nonumber\\ 
  &= \frac{1}{1+\beta}\frac{1}{\left(1+ 2 \frac{\beta}{1+\beta}vw\xi + \left(\frac{\beta}{1+\beta}\right)^2w^2v^2\right)^{3/2}} + \frac{\beta}{1+\beta}\frac{1}{\left(1- 2 \frac{1}{1+\beta}vw\xi + \left(\frac{1}{1+\beta}\right)^2w^2v^2\right)^{3/2}} \, ,
\end{align}
\end{subequations}
\end{widetext}
with $v=1/u$, $\xi = \bm n \cdot \hat{\bm e}$ and $\beta=M_\mathrm{Jup}/M_\odot\approx 10^{-3}$. 

Eq.~\eqref{eq:gN_Jup2} is similar to Eq.~\eqref{eq:gN} such that the $Q_2$ computation is identical to the one described in the previous subsection. This leads to 
\begin{subequations}\label{eq:Q_2_jup_both}
\begin{equation}\label{eq:Q2_Jup}
    Q_2' =-\frac{3a_0}{2R_M'}q'= -\frac{3a_0^{3/2}}{2\sqrt{G(M_\odot+M_\mathrm{Jup})}}q' \, ,
\end{equation}
with 
\begin{align}\label{eq:q_jup}
    q'=\frac{3}{2}\int_0^\infty \mathrm{d} v  \int_{-1}^1 d\xi &\left(\nu -1\right)\Big[ e_{\rm N}'\left(3\xi -5\xi^3\right) \\
    &\quad + \alpha' v^2\left(1-3\xi^2\right)\Big] \, . \nonumber
\end{align}
\end{subequations}
These expressions are similar to Eqs.~\eqref{eq:Q2_sun_both} with four small but notable differences: (i) the square root of the total mass $\sqrt{M_\odot+M_\mathrm{Jup}}$ now appears in the denominator of Eq.~\eqref{eq:Q2_Jup}, (ii) $e_N$ is replaced by $e_N'$ which now depends on $v$ and $\xi$, (iii) the second term in the integrand of Eq.~\eqref{eq:q_jup} has an additional $\alpha'$ factor which also depends on $v$ and $\xi$, and (iv) the $\nu$ term is now $\nu=\nu\left[\sqrt{ e_{\rm N}'^2+\alpha'^2v^4+2 \alpha'\tilde e_{\rm N}'v^2\xi}\right]$. Numerically integrating this for various interpolating functions and various values of the Galactic external field, we always find a relative difference between the cases with and without Jupiter of the order of 0.05 \%, in agreement with our intuitive expectation.

Indeed, the above expressions are exact, but since both $w\approx 6.8\times 10^{-4}$ and $\beta \approx 10^{-3}$ are actually small parameters, Eqs.~\eqref{eq:tilde_e_alpha} can also be expanded as
\begin{subequations}
\begin{align}
     e_N'&= e_N +  3 \frac{\beta}{(1+\beta)^2} v^4w^2+\mathcal O(\beta w^3)\, , \\
     \alpha' & = 1 + \frac{15}{2} \frac{\beta}{(1+\beta)^2}v^2w^2\xi^2 + \mathcal O(\beta w^3) \, . 
\end{align}
\end{subequations}
Note that the aspherical phantom density is significant for values of $\bm x$ of the order of the MOND radius, see Fig.~\ref{fig:phantom}. This means that the value of the integral from Eq.~(\ref{eq:q_jup}) depends mainly on the values of the integrand when $v=1/u=R_M'/\left|\bm x\right|\sim 1$, which means that $ e_N'=e_N+\mathcal O\left(10^{-9}\right)$ and $ \alpha' =1+\mathcal O\left(10^{-9}\right)$ as well, meaning that $q'=q +  \mathcal O \left(10^{-9}\right)$. With a very good approximation, we can then write, as intuitively expected,
\begin{align}
Q_2' \simeq
&\sqrt{\frac{M_\odot}{M_\odot+M_\mathrm{Jup}}} Q_2 \, .
\end{align}
\end{document}